# Intrinsic Grain Boundary Shear Coupling Tensor

Xinyuan Song[†], Liang Yang[§,*], Chuang Deng[†,*]

§ School of Aeronautical Manufacturing Engineering, Nanchang Hangkong University, Nanchang 330063, China

† Department of Mechanical Engineering, University of Manitoba, Winnipeg, MB R3T 2N2, Canada

* Corresponding authors: Chuang.Deng@umanitoba.ca (C. Deng); l.yang@nchu.edu.cn (L. Yang)

**Abstract**:

Grain boundary (GB) migration stands as a linchpin process governing microstructural evolution in polycrystalline materials. Over the past decade, the concept of shear coupling, quantified through the shear coupling factor, has transformed our understanding and driven the development of theoretical frameworks for unifying GB behaviors. In this study, we introduced a novel concept of shear coupling strength designed to overcome the limitations of the conventional shear coupling factor, notably its deficiency in conveying "coupling" information. The shear coupling tensor formed by the shear coupling strengths characterizes intrinsic shear coupling properties across diverse GBs and reveals complex dynamics within the GB mobility tensor. The molecular dynamics simulation confirms the symmetry of the GB mobility tensor. This symmetry is inherently built into the shear coupling strength, aligning with an assumption made in previous studies. Additionally, an efficient methodology has been developed for streamlined extraction of both shear coupling and GB mobility tensors from atomistic simulations. This advancement holds the potential to sample GB behavior across extensive datasets, significantly enhancing our ability to predict structure-property relationships within the expansive 5-parameter space of GBs.

# 1 Introduction

Grain boundary (GB) migration serves as a pivotal factor in the microstructural evolution of polycrystalline materials. Comprehending its underlying mechanisms is crucial for the targeted manipulation of both mechanical and thermal properties in these materials[1–5]. Despite over eight decades of rigorous research across experimental[6–13], theoretical[14–16], and computational[17–26] domains, reliably predicting GB migration remains a considerable challenge for the materials science community due to the inherent complexities of GB behavior and the absence of a unifying theoretical framework.

Over the past decade, the discovery and study of shear coupling effect has been transformative in refining our understanding of GB behavior. This effect, which entails a synchronized movement of the GB perpendicular to its plane as well as the relative displacement of adjacent grains (henceforth termed as GB shear migration or GB migration in y or z directions for the models in the current study), has introduced quantitative metrics for real-time observation and measurement of dynamic GB processes. This innovation has not only catalyzed the development of advanced in-situ experimental techniques[10,12,27] but also contributed to the development of comprehensive disconnection theory[28–30] that aim to unify diverse GB behaviors under a single theoretical framework.

Traditionally, the shear coupling is thought to be a geometrically related property, intimately tied to the crystallographic parameters of the GB[14,18,19,30–33]. And it is often quantified by the shear coupling factor, denoted as $\beta$, which is the ratio of GB shear displacement or velocity relative to that in the normal direction, i.e., $\beta = Z(t)/X(t)$ or $v_z/v_x$. Expanding upon the foundation of disconnection theory, Srolovitz et al. [24,29] comprehensively enumerate the shear coupling across all conceivable disconnection modes in symmetric tilt GBs and introduce a method to assess the weighted $\beta$. Further, Admal et al. [33] have broadened this approach to encompass a wider range of general GBs. However, the conventional $\beta$ only reflects the shear magnitude but overlooks the interplay or "coupling" between the displacement in two directions. For example, as illustrated in Fig. 1, given the same transition from state *a* to *c*, the GB migration can be realized either by a shear-coupled migration or by a combination of independent shear and normal migrations. The overall (or "average") $\beta$ in the two processes are the same but their shear coupling dynamics are completely different. This characteristic makes the prediction of driving force-dependent shear coupling challenging. Recent research by Chen et al.[24] demonstrates that $\beta$ within a GB can exhibit considerable variation contingent on the specific type of driving force involved, such as energy jump $\varphi$ or shear stress $\tau$. Notably, when temperatures are near the melting point, the direction of GB migration can be dictated by the orientation of the driving force[18,34], resulting in $\beta$ approaching 0 under $\varphi$ or infinity under $\tau$. Such observation has led to questions about the nature of GB shear coupling as a GB property, thereby challenging the predictability of GB behavior.

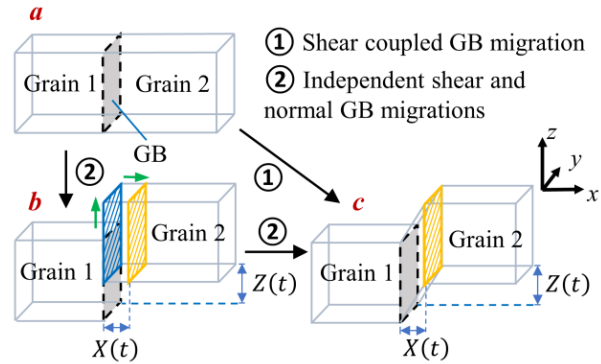

Figure 1 Schematics of two different GB migration processes with the same apparent shear coupling factor $\beta$

In this study, we introduce a shear coupling tensor, $S$, composed by the newly defined shear coupling strengths, not only quantifies the ratio of GB migration velocity or displacement in different directions, akin to $\beta$, but also factors in the degree of correlation between the GB motion across these directions, an aspect overlooked by $\beta$. This refined perspective establishes shear coupling as an intrinsic GB property, enables a mathematical confirmation of the symmetry of GB mobility tensor and illuminates the core relationship between shear coupling and intrinsic GB mobility across different directions. Furthermore, a method is developed to conveniently compute shear coupling and mobility tensors directly from the interface random walk simulations.

## 2 Methods

### 2.1 Atomistic modeling

All the models used in this study are from Olmsted 388 Ni GBs database[35]. Before the simulations, the models were first expanded at various temperatures, according to the corresponding thermal expansion coefficient, and subsequently equilibrated under the isothermal-isobaric ensemble (NPT) for 10 ps. This was followed by a brief annealing period of 5 ps under the microcanonical ensemble (NVE) with a Berendsen thermostat (Berendsen NVT ensemble)[36].

### 2.2 Interface Random Walk Simulations

After the model reached the equilibriums at the target temperatures, the Berendsen thermostat was removed, and GBs were subjected to random fluctuations under the NVE ensemble for a duration of 5 ns. The simulation at each temperature was performed 20 times, using different random seeds for the initial velocity distribution. During the post-simulation data processing, each 5 ns simulation was segmented into 10 parts of 500 ps each, thereby yielding 200 sets of independent simulation data for each temperature. We utilized the order parameter as defined in Ref.[37] to track the GB displacement in the $x$ direction, $X(t)$, while the GB shear displacements in the $z$ direction, $Z(t)$, were computed based on the relative shear movement of the center of mass of a thin slab, roughly 1 nm thick, situated at both ends of the model in the $x$ direction, as illustrated as blue parts in Fig. 2.

### 2.3 GB Migration Simulations with External Driving Forces

The energy-conserving orientational (ECO) synthetic driving force method[37,38] was applied as the normal driving force $\varphi$. For the shear stress $\tau$, two opposing shear forces were applied to the thin slabs at the ends of the model in $x$ direction. Simulations were conducted under the Berendsen NVT ensemble until the GB reached one end of the model, taking a maximum of 5 ns. Each simulation was replicated 20 times, utilizing different random seeds for the initial velocity distribution. The velocity of GB migration was computed by fitting the slope of the cumulative GB displacement *vs. t* line. The elements in GB mobility tensor at finite driving forces can then be calculated from

$$\begin{bmatrix} v_x \\ v_y \\ v_z \end{bmatrix} = \begin{bmatrix} M_{xx} & M_{xy} & M_{xz} \\ M_{yx} & M_{yy} & M_{yz} \\ M_{zx} & M_{zy} & M_{zz} \end{bmatrix} \begin{bmatrix} \varphi \\ \tau_y \\ \tau_z \end{bmatrix} \quad (1)$$

It should be noted that Eq. 1 hinges on a critical assumption: the mobility *M* remains unaffected by external driving forces and the velocity *v* changes linearly with these forces. This assumption is generally valid for small driving forces in molecular dynamics (MD) simulations. The subsequent comparison of mobility values calculated using Eq. 1 with those obtained from interface random walk simulations in which the driving force is zero will further validate this assumption.

## 3 Results

We began our exploration with the Σ15 (2 1 1) Ni GB (designated as P14 in the Olmsted database[35]), as shown in Fig. 2. This GB exhibits negligible mobility in the *y* direction, with the shear coupling effect manifesting predominantly in the *x-z* plane, as illustrated later in Fig. 3, offering an ideal starting study case. According to the recently introduced concept of GB mobility tensor[39], the kinetic equation governing the GB migration can be expressed as[29,30]

$$\begin{bmatrix} v_x \\ v_z \end{bmatrix} = \begin{bmatrix} M_{xx} & M_{xz} \\ M_{zx} & M_{zz} \end{bmatrix} \begin{bmatrix} \varphi \\ \tau \end{bmatrix} = \begin{bmatrix} M_{xx} & M_{zz}/\beta_\tau \\ \beta_\varphi M_{xx} & M_{zz} \end{bmatrix} \begin{bmatrix} \varphi \\ \tau \end{bmatrix} \quad (2)$$

Here $\beta_\varphi$ and $\beta_\tau$ are the shear coupling factors computed under energy jump $\varphi$ and shear stress $\tau$, respectively, as utilized in prior studies[18,24,30,34,35].

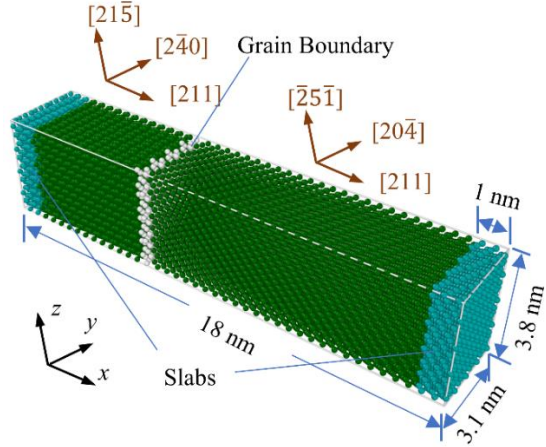

Figure 2 The atomistic model of Ni Σ15 (2 1 1) GB

To prevent alterations in the GB migration mechanism and mobility that can result from unrealistic large driving forces in MD simulations[40–42], the interface random walk method[43] was employed to extract the GB mobility as well as the shear coupling within it. In this method, no specific type of driving force is applied, and the computed GB mobilities are close to that in experimental conditions. Simulations were conducted using the LAMMPS package[44], utilizing an embedded atom method potential specifically for Ni[45]. Fig. 3 illustrates the outcomes of the interface random walk simulation for Σ15 (2 1 1) Ni GB at 200 K. The data show a significant and random displacement of the GB in both the *x* and *z* directions (with the average position remaining at 0), whereas the displacement in the *y* direction remained stable, implying a close-to-zero mobility along this axis ($M_{yy} \approx 0$).

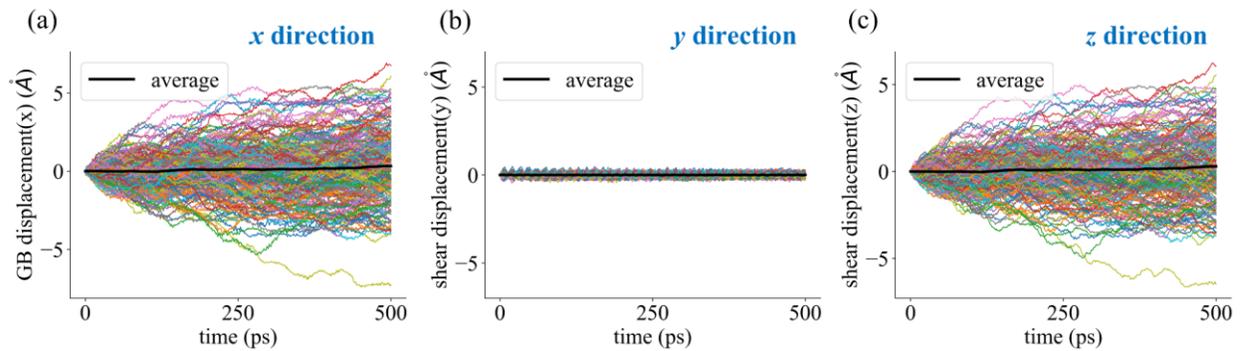

Figure 3 The fluctuation of the average position of Ni Σ15 (2 1 1) GB in different directions at 200 K.

The interface random walk method is based on the observation that, the average GB displacement across a substantial number of independent simulations at any specific time $t = j\Delta t$ is approximately 0, as shown in Fig. 3, while the variance of GB displacement increases linearly with the $t$ multiplied by the diffusion coefficient $D$[43,46,47], expressed as

$$\sigma_j^2 = \frac{\sum_{i=1}^{m} X_i^2(j\Delta t)}{m-1} = Dj\Delta t \qquad (3)$$

where $m$ is the total number of independent simulations, $\Delta t$ is the timestep, and $X_i(j\Delta t)$ is the position of the GB at time $j\Delta t$ in simulation $i$. It should be emphasized that in this context, $D$ is derived from the mean square displacement (MSD) of the average normal migration across the entirety of the GB plane, rather than the individual atoms. This value is directly related to the normal mobility, denoted as $M_{xx}$, and the area, $A$, of the GB, which can be mathematically expressed as[43]:

$$D = \frac{2M_{xx}k_B T}{A} \qquad (4)$$

where $k_B$ is Boltzmann constant, and $T$ is temperature. Acquiring a meaningful variance $\sigma_j^2$ demands that $m$ be sufficiently large. This requirement poses a significant challenge in terms of computational power and restricts its applicability for sampling GB mobilities over an extensive dataset of GBs. To overcome this constraint, we introduce the Fast Adapted Interface Random Walk (FAIRWalk) method which could enhance statistical accuracy by orders of magnitude without necessitating additional simulations. The method relies on the crucial observation that the variance of GB displacement accumulated throughout the entire process also encapsulates the essential information for extracting GB mobility. For instance, the variance of GB displacement in the normal direction, accumulated from $t = 0$ to $n\Delta t$ can be computed as:

$$\sigma_x^2 = \frac{\sum_{j=1}^{n}\sum_{i=1}^{m} X_i^2(j\Delta t)}{nm-1} = \frac{\sum_{j=1}^{n}(m-1)Dj\Delta t}{nm-1} = \frac{(m-1)D\Delta t}{nm-1} \cdot \frac{n(n+1)}{2} \qquad (5)$$

When both $m$ and $n$ are sufficiently large, we can obtain:

$$\sigma_x^2 = \frac{D}{2} n \Delta t \tag{6}$$

To evaluate the performance of the FAIRWalk method against the current methodologies, we plotted the $\sigma_x^2$ vs. $t$ curve obtained from Eq. 5, and $\sigma_j^2$ vs. $t$ curves derived from both Eq. 3 and the adapted interface random walk(AIRWalk) method proposed by Deng and Schuh[47], which involves fitting the distribution of GB displacement from independent simulations at a certain moment to an error function to determine $\sigma_j^2$. Fig. 4d illustrates that the $\sigma_x^2$ vs. $t$ curve demonstrates a significantly stronger linear relationship, characterized by a slope equal to $D/2$. This approach further increases precision by systematically integrating all extant data points in the computation of $\sigma_x^2$. Subsequently, the GB mobility can be computed using the Eq. 4:

$$M_{xx} = \frac{DA}{2k_B T} = \frac{A}{k_B T t_0} \sigma_x^2 \tag{7}$$

Here, $t_0 = n\Delta t$ is the total simulation time. The FAIRWalk method can mitigate the need for substantial computational power to generate the large number of data points necessary for an unbiased statistical analysis, as required by conventional interface random walk methods. For example, in the case shown in Fig. 4d with $m = 200$, $t_0 = 500$ $ps$, $\Delta t = 1$ ps, and $n = t_0/\Delta t = 500$, the computation of $\sigma_x^2$ and subsequently $D$ and $M$ in the FAIRWalk method is based on $nm = 100{,}000$ independent data points, which could be significantly reduced in practice while still achieving results with a similar level of confidence. The efficiency and capability of FAIRWalk in extracting mobilities for various types of GB will be explored further in a subsequent study.

The FAIRWalk method can be extended to the shear direction by substituting $\sigma_x^2$ with $\sigma_z^2$[46]. Figs. 4e-f demonstrate that the computed GB normal and shear mobilities are consistent with $M_{xx}$ and $M_{zz}$ in the GB mobility tensor (Eq. 2) when a small synthetic energy jump $\varphi$[37,38] or a shear stress $\tau$ is applied. This also corroborates the applicability of Eq. 1 under small driving forces. Importantly, this range should encompass the typical experimental driving forces of $10^2$ - $10^6$ Pa[42].

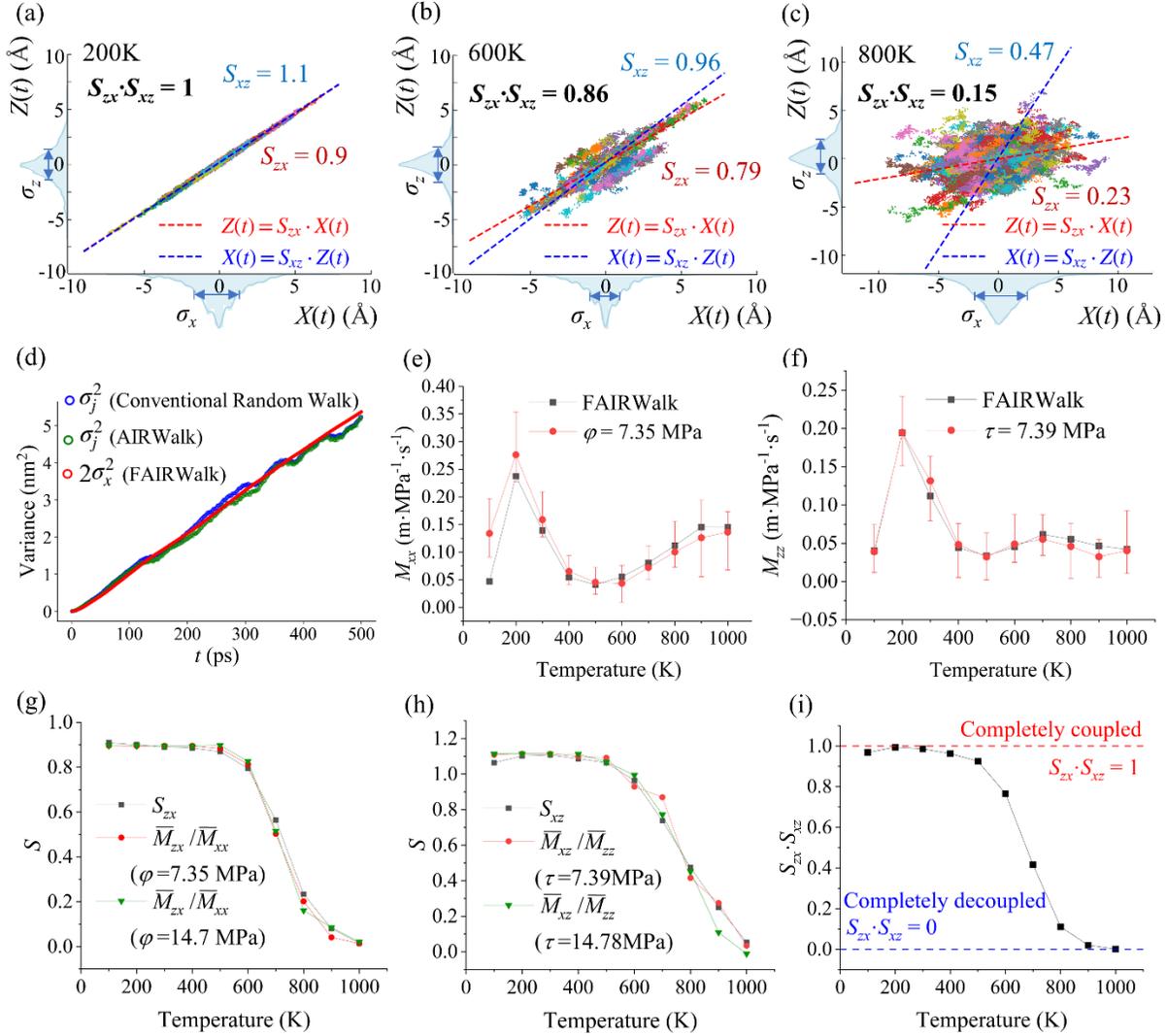

Figure 4 (a-c) Illustration of shear coupling strength $S$ and the variance $\sigma^2$ from random walk simulation at different temperatures (dots with different colors represent the data from simulations with different initial velocity distribution). (d) The comparison between $\sigma_j^2$ in Eq. 3 from previous random walk methods and $2\sigma_x^2$ in Eq. 5 $vs.$ $t$. (e-f) Comparison of $M_{xx}$ and $M_{zz}$ computed by FAIRWalk method with those calculated from simulations with external driving forces. (g-h) Comparison of $S$ with the shear coupling relation between elements in the GB mobility tensor from simulations with external driving forces. (the mobility with an overbar $\bar{M}$ indicates the mean value) (i) The evolution of $S_{zx} \cdot S_{xz}$ with temperature as an indicator for shear coupling status.

Furthermore, defining the shear coupling factor, either as $\beta = Z(t)/X(t)$ or within Chen et al.'s GB mobility tensor (e.g., $\beta_\varphi = M_{zx}/M_{xx}$ or $\beta_\tau = M_{zz}/M_{xz}$)[24,39], renders the extraction of shear coupling information from the interface random walk simulation results challenging, because the average GB displacement in *x* and *z* directions would be both zero. Extracting *β* as the slope from a linear fitting of *Z(t)* vs. *X(t)*, as one may attempt, would lead to an erroneous relation that $\beta_\varphi$ is always equal to $\beta_\tau$. To address this issue, we applied the least squares linear regression[48] to fit both the *Z(t)* vs. *X(t)* and *X(t)* vs. *Z(t)* data, as illustrated in Figs. 4a-c. The fitted slope can be mathematically expressed as (see Supplemental Section S1 for details)

$$S_{xz} = \rho \frac{\sigma_x}{\sigma_z}, S_{zx} = \rho \frac{\sigma_z}{\sigma_x} \tag{8}$$

Here, $\rho$ is the Pearson correlation coefficient[48], ranging from -1 to 1, serves as a standard measurement of the linear correlation strength or "coupling" between *X(t)* and *Z(t)*. And the ratio of the variabilities of *X(t)* and *Z(t)*, namely $\sigma_z/\sigma_x$ or $\sigma_x/\sigma_z$, adjusts the scale of the relationship, which could also be interpreted as the "shear" magnitude. Integrating these components, we define $S_{zx}$ and $S_{xz}$ as the GB shear coupling strengths, which quantify the extent to which GB displacement in one direction directly results from the displacement in the other direction. The newly defined shear coupling strength not only inherently complies with the Onsager relation for the ***M*** tensor but also yields the same analytical form of the entire ***M*** tensor as that based on the mathematical derivation of the GB kinetics equation for the 3D interface random walk. The latter will be published elsewhere in a separate study. Additionally, the shear coupling strength also holds the statistical significance. According to Han et al.[30], GB migration is mediated by various disconnection modes (See Supplemental Fig. S2 for possible disconnection modes in Σ15 (2 1 1) GB), with the activation of these modes adhering to the Boltzmann distribution[29]. In a long run, there exists a statistical shear coupling intrinsic to the GB, and the fitted slope (*S*) between the displacements in two directions obtained through linear regression serves as a representation of this GB property.

Figs. 4g and h demonstrate that $S_{zx}$ and $S_{xz}$, derived from the interface random walk simulations, can accurately capture the shear coupling information between the elements in the GB mobility tensor, and this shear coupling remains constant when the driving force is small. These observations support our hypothesis that shear coupling, akin to GB mobility, ought to be recognized as an intrinsic property of GB. It is worth noting that although the driving forces applied in our tests (7.3 and 14.7 MPa) are almost the minimum we can utilize in MD simulations to observe apparent GB migration within 5 ns, they are nearly the maximum possible driving forces in experimental scenarios (typically $10^2 - 10^6$ Pa[42]). Consequently, this finding regarding the intrinsic property of shear coupling is expected to be applicable in most realistic situations. Therefore, we can reformulate the GB mobility tensor and the kinetic equation as:

$$\begin{bmatrix} v_x \\ v_z \end{bmatrix} = \begin{bmatrix} 1 & S_{xz} \\ S_{zx} & 1 \end{bmatrix} \begin{bmatrix} M_{xx} & \square \\ \square & M_{zz} \end{bmatrix} \begin{bmatrix} \varphi \\ \tau \end{bmatrix} = \boldsymbol{S}\boldsymbol{M}_p \begin{bmatrix} \varphi \\ \tau \end{bmatrix} \qquad (9)$$

Here, $\boldsymbol{M}_p$ represents the tensor comprising the principal GB mobilities, while $\boldsymbol{S}$ defined as the GB shear coupling tensor. The interrelation among the elements in $\boldsymbol{S}$ is given by

$$S_{xz} \cdot S_{zx} = \rho^2 \qquad (10)$$

where $\rho^2$, ranging between 0 and 1, serves as a normalized measure of the linear relationship between $X(t)$ and $Z(t)$, effectively characterizing the shear coupling status of GB migration. As shown in Fig. 4i, $S_{zx} \cdot S_{xz}$ shifts from 1 to 0 with the increase in temperature. This shift indicates a gradual decoupling transition from completely coupled GB migration to entirely independent GB normal and shear migrations, devoid of shear coupling. The disconnection theory[9,26,30,49] provides a plausible explanation for this phenomenon. At low temperatures, only the mode requiring the minimum activation energy can be activated, leading to a fixed ratio of GB normal and shear migration and a perfect linear correlation between the two (Fig. 4a). However, as the temperature rises, multiple modes can be activated [24,30]. This multiplicity implies that a specific GB displacement in one direction could correspond to several displacements in other directions, thereby diminishing their linear correlation (Figs. 4b and c). Ultimately, when the temperature nears the melting point, the GB migrations in various directions become independent, as the linear correlation between them essentially vanishes. At this stage, GB migration can occur randomly in different directions. Although each migration combination yields a specific shear coupling factor $\beta$, there is, in fact, no actual shear coupling between migrations in different directions, as evidenced by zero values of $S_{zx}$, $S_{xz}$ or $S_{zx} \cdot S_{xz}$.

Furthermore, we can deduce from Eqs. 7 and 8 that $\frac{M_{xx}}{M_{zz}} = \frac{\sigma_x^2}{\sigma_z^2} = \frac{S_{xz}}{S_{zx}}$, and with Eq. 9, we arrive at

$$M_{zx} = M_{xz} = \frac{A}{k_B T t_0} \rho \sigma_x \sigma_z \tag{11}$$

Eq. 11 provides a validation of Chen et al.'s[39] assumption that Onsager symmetry is preserved within the GB mobility tensor. Our MD simulation results, depicted in Fig. 5, confirm this conclusion.

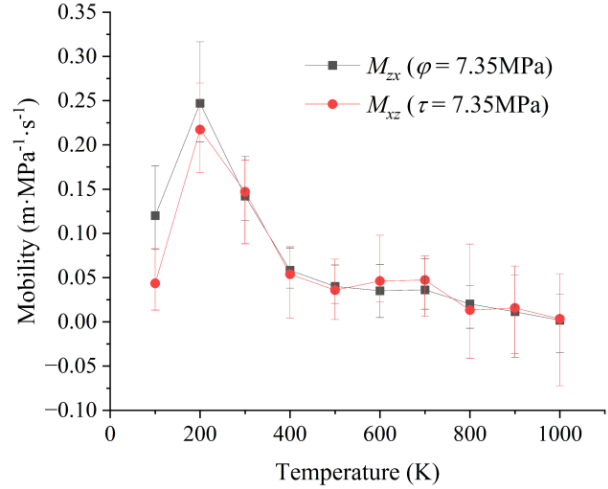

Figure 5 Comparison between $M_{zx}$ and $M_{xz}$ of Σ15 (2 1 1) Ni GB. The $M_{zx}$ and $M_{xz}$ are computed from simulations with different types of driving forces ($\varphi$ and $\tau$), respectively.

Consequently, Eq. 2 can be reformulated as

$$\begin{bmatrix} v_x \\ v_z \end{bmatrix} = M_{xx} \begin{bmatrix} 1 & S_{zx} \\ S_{zx} & S_{zx}/S_{xz} \end{bmatrix} \begin{bmatrix} \varphi \\ \tau \end{bmatrix} \tag{12}$$

Eq. 12 unveils an intrinsic connection among the components of ***M***, illustrating that this connection is encapsulated within a tensor composed exclusively of shear coupling strengths. This insight emphasizes the nature of shear coupling as an inherent GB property: since GB mobility is widely regarded as an intrinsic feature of GB under typical experimental conditions, the same consideration should be extended to GB shear coupling.

## 4 Discussions

### *4.1 Driving force dependent shear coupling factor β and fixed shear coupling tensor S*

Suspicions about the shear coupling as a property of GB primarily stem from the driving force dependency observed in the traditional shear coupling factor $\beta$ [24]. Upon examining Eqs. 8 and 19, it becomes evident that $\beta$ under $\varphi$ and $\beta$ under $\tau$, represented by $M_{zx}/M_{xx}$ and $M_{zz}/M_{xz}$ respectively, do not share identical physical interpretations. This discrepancy elucidates the contrasting variation trends of $\beta$ under different driving forces as noted in a prior study[24], as illustrated in Fig. 6a. It further suggests that $\beta$ will be inadequate for analyzing GB shear coupling under complex driving forces encountered in real polycrystalline materials. For example, as illustrated in Figs. 6b and c, the apparent $\beta$ of Σ15 (2 1 1) Ni GB under a composite driving force of $\varphi$ = 1.47 MPa and $\tau$ = 7.39 MPa at 800 K is $v_z/v_x$ = 1.32, and when $\varphi$ increases to 7.35 MPa, the apparent $\beta$ shifts to 0.54. This significant driving force dependency, even for relatively small forces, reinforces Chen et al.'s claim[24] that the apparent $\beta$, as commonly reported in previous studies, is not an intrinsic measure of GB shear coupling. In contrast, by utilizing the same $S_{xz}$ and $S_{zx}$ at 800 K extracted from random walk simulation (as illustrated in Fig. 4c), the predicted $v_z/v_x$ ratios from Eq. 12 for the specified composite driving forces above are 1.25 and 0.58, respectively. These values align closely with the ratios of 1.32 and 0.54 obtained directly from atomistic simulations. This alignment underscores the fact that under typical experimental conditions when no change in migration mechanism occurs, although the apparent $\beta$ is sensitive to the driving force, $S$ remains a fixed GB property. Consequently, Eq. 12 serves as an accurate means to predict the apparent $\beta$.

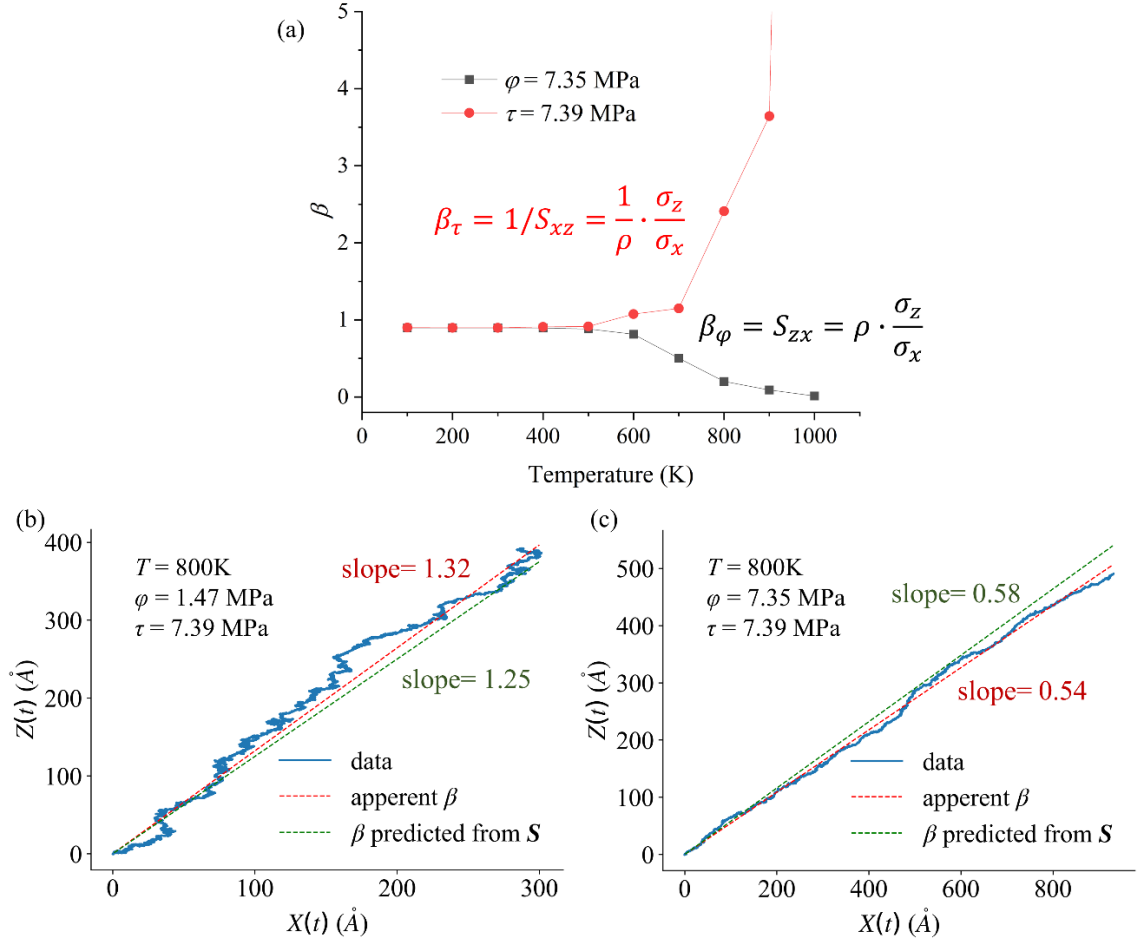

Figure 6 (a) The expression of apparent $\beta$ by $S$ under single driving forces. (b, c) The comparison between the apparent $\beta$ and the $\beta$ predicted from $S$ under composite driving forces of (b) $\varphi = 1.47$ MPa, $\tau = 7.39$ MPa and (c) $\varphi = 7.35$ MPa, $\tau = 7.39$ MPa. $X(t)$ and $Z(t)$ are accumulated displacements obtained from twenty 5 ns simulations.

## 4.2 The asymmetry of shear coupling strengths

An intriguing feature noted in Figs. 4(a-c) is the asymmetrical nature of the shear coupling strengths (represented by the slopes fitted by the linear regression), i.e. $S_{zx} \neq 1/S_{xz}$. This asymmetry is elucidated in Eq. 8, where the linear correlation coefficient $\rho$ remains constant for any two given variables, irrespective of their roles as predictor or predicted value. However, the scale factor, represented by $\sigma_z/\sigma_x$ or $\sigma_x/\sigma_z$, is symmetric when the predictor role is switched between the two variables. Symmetry in $S_{zx}$ and $S_{xz}$ is only observed when the $X(t)$ and $Z(t)$ are fully coupled, i.e. when $\rho = 1$, leading to $S_{zx} = \sigma_z/\sigma_x = 1/S_{xz}$.

This relationship reveals the fact that GB shear coupling, when GB migration is not fully coupled, experiences asymmetric effects from different types of driving force. The asymmetry likely stems from how different types of driving force preferentially activate different groups of disconnection modes ($b_i$, $h_i$) during GB migration. This phenomenon is consistent with the previously reported dilemma that as the driving force approaches zero, the apparent $\beta$ converges to different values depending on the type of driving force, i.e. $\beta_{\varphi \to 0} \neq \beta_{\tau \to 0}$ [24]. And this discrepancy is only absent at low temperatures where the GB displacements in different directions are fully coupled.

*4.3 Three-dimensional shear coupling tensor S and GB mobility tensor M*

For a GB that exhibits shear coupling in all three directions, such as the Σ315 (10 6 2) Ni GB (P182 in the Olmsted database[35]), the cumulated GB displacements in random walk simulations can be plotted in a 3D graph, as illustrated in Fig. 7. This single plot encapsulates all the crucial data needed to calculate both the GB mobility and shear coupling tensors. For example, the diagonal GB mobilities ($M_{xx}$, $M_{yy}$ and $M_{zz}$) can be calculated using Eqs. 5 and 7, based on the variance of GB displacements in each direction. Meanwhile, the shear coupling strengths can be extracted by linearly fitting the data points within the respective axis planes.

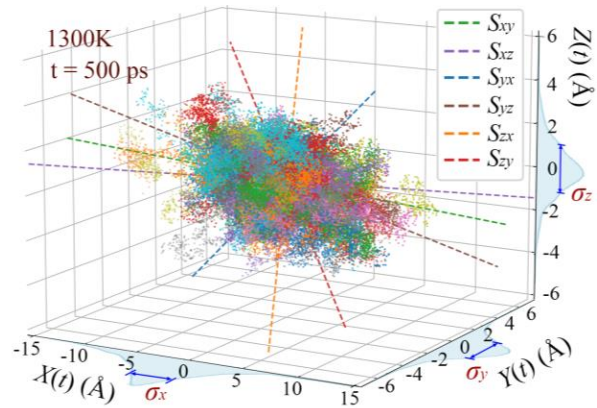

Figure 7 3D plot of the cumulated displacements in a Ni Σ315 (10 6 2) GB at 1300 K, from which the shear coupling strength $S$ and variance $\sigma^2$ in each direction can be extracted (dots with different colors represent the data from simulations with different initial velocity distributions).

Finally, the three-dimensional GB mobility tensor $\boldsymbol{M}$ can be reformulated into a diagonal tensor containing only the principal mobilities (i.e., mobilities when shear coupling is not present), multiplied by the shear coupling tensor $\boldsymbol{S}$. And the kinetic equation for GB migration can be expressed by

$$\begin{bmatrix} v_x \\ v_y \\ v_z \end{bmatrix} = \boldsymbol{M} \begin{bmatrix} \varphi \\ \tau_y \\ \tau_z \end{bmatrix} = \boldsymbol{S} \begin{bmatrix} M_{xx} & & \\ & M_{yy} & \\ & & M_{zz} \end{bmatrix} \begin{bmatrix} \varphi \\ \tau_y \\ \tau_z \end{bmatrix} \quad (13)$$

where

$$\boldsymbol{S} = \begin{bmatrix} 1 & S_{xy} & S_{xz} \\ S_{yx} & 1 & S_{yz} \\ S_{zx} & S_{zy} & 1 \end{bmatrix} \quad (14)$$

By extending Eq. 7 to all principal mobilities, the $\boldsymbol{M}$ tensor can also be expressed as

$$\boldsymbol{M} = \frac{A}{k_B T t_0} \begin{bmatrix} \sigma_x^2 & \rho_{xy}\sigma_x\sigma_y & \rho_{xz}\sigma_x\sigma_z \\ \rho_{xy}\sigma_x\sigma_y & \sigma_y^2 & \rho_{yz}\sigma_y\sigma_z \\ \rho_{xz}\sigma_x\sigma_z & \rho_{yz}\sigma_y\sigma_z & \sigma_z^2 \end{bmatrix} \quad (15)$$

highlighting the inherent symmetry of $\boldsymbol{M}$ tensor. It also elucidates the relational dynamics among the elements in $\boldsymbol{S}$ tensor: $S_{xy}S_{yz}S_{zx} = S_{yx}S_{zy}S_{xz} = \rho_{xy}\rho_{yz}\rho_{xz}$. See Supplemental Section S3 for verification of Eq. 13 in the Σ315 (10 6 2) Ni GB based on atomistic simulations at temperatures between 500 – 1400 K.

**5 Conclusion**

Understanding GB behavior is vital for predicting microstructural transformations in materials subjected to deformation or heat treatments, providing an essential foundation for the progression of future industrial applications[50,51]. In this study, we introduce a novel concept of shear-coupling strength and its associated tensor $\boldsymbol{S}$ which surpass the limitations inherent in the widely employed conventional $\beta$ factor and offer a more accurate depiction of GB shear coupling properties. Utilizing this advancement, we can draw several key conclusions:

1. Intrinsic nature of shear coupling tensor $S$. Unlike the conventional $\beta$ factor, which is sensitive to external influences, the newly introduced shear coupling tensor $S$ is an inherent attribute of GBs. This significant advancement bridges a crucial gap and enhances the predictive capacity regarding the structure-property interplay within the extensive 5-parameter space of GBs.

2. Temperature-dependent decoupling transition. Our investigation into the tensor $S$ has unveiled a progressive shift from complete shear coupling to a state of no coupling with rising temperatures. This finding sheds light on the thermal sensitivity of GB behavior.

3. Interconnection between GB mobility and shear coupling. We have uncovered that GB mobility and shear coupling are fundamentally linked, with the relationship between all elements in $M$ can be exclusively expressed by shear coupling strengths. Moreover, the symmetry of $M$, aligning with the Onsager relation, is inherently built into the shear coupling strength and confirmed by our MD simulation results.

4. FAIRWalk method. The FAIRWalk method, devised specifically for this study, facilitates the efficient and precise extraction of $M$ and $S$ tensors from the interface random walk simulations.

The elucidation of GB shear coupling and mobility tensors through this study not only deepens our understanding of GB dynamics in complex composite stress environments but also aids in interpreting microstructural transformations in polycrystalline materials. The symmetry properties inherent to $M$ tensor merit further exploration, presenting promising opportunities for future GB engineering endeavors.

**Acknowledgment**

The authors thank Dr. David L Olmsted for sharing the 388 Ni GB structure database. This research was supported by NSERC Discovery Grant (RGPIN-2019-05834), Canada, and the use of computing resources provided by Digital Research Alliance of Canada. X.S. also acknowledges financial support from UMGF, the University of Manitoba Graduate Fellowship. During the preparation of this manuscript the authors used ChatGPT to improve its readability. After using this tool, the authors reviewed and edited the manuscript as needed and take full responsibility for the content of the publication.

# Supplementary materials

# for

# Intrinsic Grain Boundary Shear Coupling Tensor

Xinyuan Song[†], Liang Yang[§,*], Chuang Deng[†,*]

§ School of Aeronautical Manufacturing Engineering, Nanchang Hangkong University, Nanchang 330063, China

† Department of Mechanical Engineering, University of Manitoba, Winnipeg, MB R3T 2N2, Canada

* Corresponding authors: Chuang.Deng@umanitoba.ca (C. Deng); l.yang@nchu.edu.cn (L. Yang)

## S1. Definition of shear coupling strength from linear regression

Given a set of data points: $(x_1, z_1), (x_2, z_2) \dots (x_n, z_n)$, the simple linear regression will fit a line of the form $z = Sx + b$ with the minimum residuals sum of squared (RSS) which is given by[1]

$$RSS = \sum_{i=1}^{n}[z_i - (Sx_i + b)]^2 \qquad (S1)$$

Taking the derivative of RSS with respect to $S$, and setting it to zero, we end up with

$$S = \frac{n\sum x_i z_i - \sum x_i \sum z_i}{n\sum x_i^2 - (\sum x_i)^2} \qquad (S2)$$

Dividing the numerator and the denominator of the right-hand side by $n^2$, we get

$$S = \frac{\sum x_i z_i /n - (\sum x_i /n)(\sum z_i /n)}{\sum x_i^2 /n - (\sum x_i /n)^2} = \frac{\text{Cov}(x,z)}{\sigma_x^2} \qquad (S3)$$

The Pearson correlation coefficient is defined as

$$\rho = \frac{Cov(X,Z)}{\sigma_x \sigma_z} \tag{S4}$$

Insert Eq. S4 into Eq. S3, we get the final form of the shear coupling strength as:

$$S = \rho \frac{\sigma_z}{\sigma_x} \tag{S5}$$

## S2. Dichromatic analysis of Σ15 (2 1 1) Ni grain boundary (GB)

According to disconnection theory, grain boundary (GB) migration is mediated by disconnections, which are line defects moving within GB. These disconnections are composed of a Burgers vector *b* and height *h*, and they have been observed in both simulations[2–4] and experiments[5]. Fig. S1 illustrates some of the possible disconnection modes in the Σ15 (2 1 1) GB. Generally, the lower the values of *b* and *h*, the lower the activation energy it needs. At low temperatures, only the mode with the lowest activation energy (indicated by black arrows in Fig. S1) can be activated, leading to a shear coupling factor *β* of 0.89 for the model. This result aligns with our simulation findings (see Fig. 4g in the main article). As the temperature increases, more modes with opposing shear coupling factor signs may be activated, weakening the correlation of GB displacement in different directions. At this time, the *β* is not a fixed value.

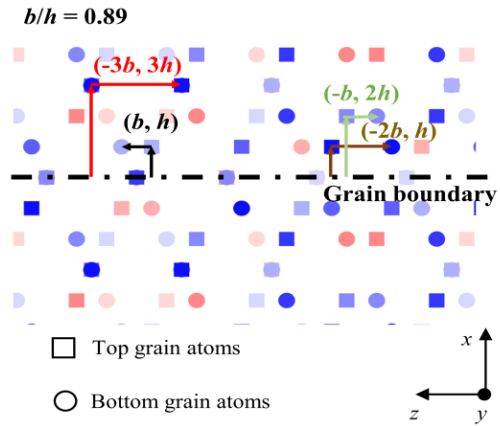

Figure S1 Dichromatic pattern of Σ15 (2 1 1) Ni GB

**S3. Verification of the GB mobility tensor M in Σ315 (10 6 2) Ni GB extracted by Fast Adapted Interface Random Walk (FAIRWalk) method.**

The classical kinetic equation for GB migration can be expressed using the GB mobility tensor as follows

$$\begin{bmatrix} v_x \\ v_y \\ v_z \end{bmatrix} = \begin{bmatrix} M_{xx} & M_{xy} & M_{xz} \\ M_{yx} & M_{yy} & M_{yz} \\ M_{zx} & M_{zy} & M_{zz} \end{bmatrix} \begin{bmatrix} \varphi \\ \tau_y \\ \tau_z \end{bmatrix} \quad (S6)$$

According to the FAIRWalk method we proposed, the diagonal elements of the GB mobility tensor $M_p$ can be expressed as

$$\boldsymbol{M_p} = \begin{bmatrix} M_{xx} & \square & \square \\ \square & M_{yy} & \square \\ \square & \square & M_{zz} \end{bmatrix} = M_{xx}\boldsymbol{S_p} \quad (S7)$$

where

$$\boldsymbol{S_p} = \begin{bmatrix} 1 & \square & \square \\ \square & S_{yx}/S_{xy} & \square \\ \square & \square & S_{zx}/S_{xz} \end{bmatrix} \quad (S8)$$

Accordingly, the complete GB mobility tensor can be expressed as

$$\boldsymbol{M} = \begin{bmatrix} M_{xx} & M_{xy} & M_{xz} \\ M_{yx} & M_{yy} & M_{yz} \\ M_{zx} & M_{zy} & M_{zz} \end{bmatrix} = \boldsymbol{S}\boldsymbol{M_p} \quad (S9)$$

with the 3-dimensional shear coupling tensor $\boldsymbol{S}$ expressed as

$$\boldsymbol{S} = \begin{bmatrix} 1 & S_{xy} & S_{xz} \\ S_{yx} & 1 & S_{yz} \\ S_{zx} & S_{zy} & 1 \end{bmatrix} \quad (S10)$$

Therefore, the kinetic equation for GB migration can be expressed by

$$\begin{bmatrix} v_x \\ v_y \\ v_z \end{bmatrix} = M_{xx} \cdot \boldsymbol{SS_p} \begin{bmatrix} \varphi \\ \tau_y \\ \tau_z \end{bmatrix} \quad (S11)$$

with

$$\boldsymbol{SS_p} = \begin{bmatrix} 1 & S_{yx} & S_{zx} \\ S_{yx} & S_{yx}/S_{xy} & S_{yz}S_{zx}/S_{xz} \\ S_{zx} & S_{zy}S_{yx}/S_{xy} & S_{zx}/S_{xz} \end{bmatrix} \quad (S12)$$

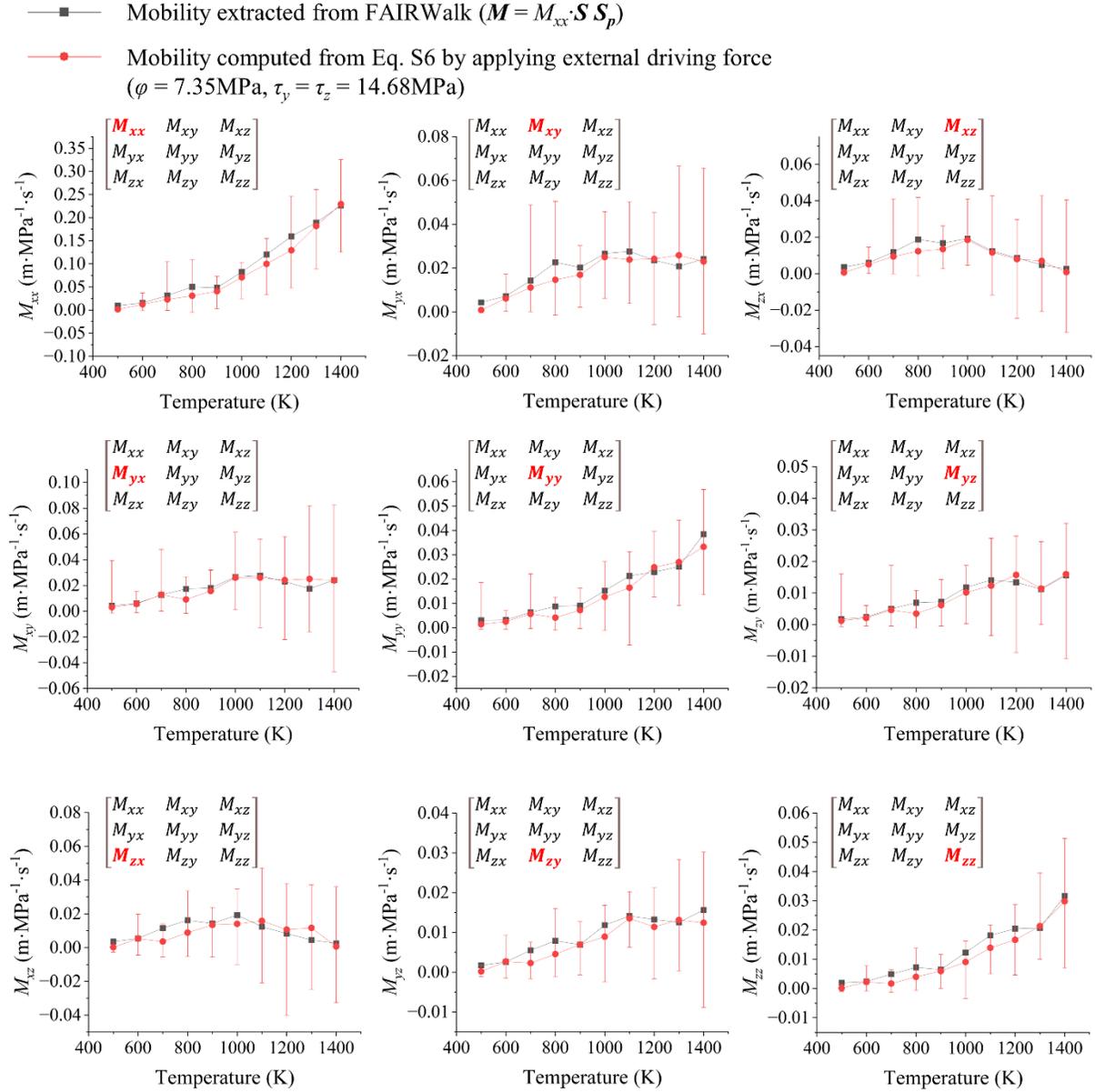

Figure S2 Comparison between GB mobility tensor of Σ315 (10 6 2) Ni GB computed by FAIRWalk method and that computed from simulations with different types of driving forces.

Fig. S2 illustrates the comparison between the GB mobility tensor calculated using the FAIRWalk method and the tensor obtained directly from simulations with external driving forces, as computed by Eq. S6. The methodologies are detailed in Section 2 in the main article. In traditional approaches, computing all elements in the GB mobility tensor requires simulations with three distinct types of driving forces. However, with the FAIRWalk method, all essential information can be gathered from interface random walk simulations. Fig. S2 emphasizes the reliability of our FAIRWalk method, achieving equivalent accuracy with fewer simulations and reduced computational costs.